\documentclass{article}%
\usepackage{amssymb}
\usepackage{amsmath}%
\usepackage{amsfonts}%
\usepackage{graphicx}

\author{Vladimir K. Petrov\thanks{ E-mail address: vkpetrov@yandex.ru}}
\title{Errors and ambiguity in transition from Fourier series to Fourier
integrals}
\date{\textit{N. N. Bogolyubov Institute for Theoretical Physics}\\
\textit{\ National Academy of Science of Ukraine}\\
\textit{\ 252143 Kiev, Ukraine. 24.11.2004}}
\begin{document}
\maketitle

\begin{abstract}
Transition from Fourier series to Fourier integrals is considered and error
introduced by ordinary substitution of integration for summing is estimated.
Ambiguity caused by transition from discrete function to continuous one is
examined and conditions under which this ambiguity does not arise are suggested.

\end{abstract}

\section{Introduction}

Quite often it appears to be appropriate to solve some physical problems at
first in a finite size box (see e.g. \cite{ilgenfritz,aoki,baig}) and to
increase the volume infinitely only at the final stage. In particular, such
approach provides a reasonable regularization. As a rule periodic border
conditions are imposed, so Fourier transform in spatial variables becomes a
discrete function (Fourier coefficients), which is assumed to convert into a
continuous one, when borders are pulled apart and period $2\pi\tau$\ is
infinitely increased. Ordinary substitution integration for summing is
acceptable for 'smooth enough' and 'fast enough' decreasing functions, but
being applied to distributions it may introduce errors which we estimate in
this paper. Moreover, a transition from a discrete function to a continuous
one may introduce ambiguity. We estimate possible ambiguity and consider
conditions under which such ambiguity doesn't arise. It should be noted that
even if the original function is smooth, after transition $\tau\rightarrow
\infty$ it may convert into a distribution.

In this paper we confine ourselves to the case of tempered distributions
$F\left(  x\right)  \in S^{\prime}$, which are defined as functionals
$\left\langle F\left(  x\right)  ,\phi\left(  x\right)  \right\rangle $ on
fast decreasing test functions $\phi\left(  x\right)  \in S$
\cite{gel-shil,vladimirov,bremermann}. Recall that $F\left(  x\right)  $ is a
tempered distribution, if and only if it is a finite order derivative of some
continuous tempered function $G\left(  x\right)  $, i.e. $F\left(  x\right)
=G^{\left(  n\right)  }\left(  x\right)  $, $n<\infty$. Function $G\left(
x\right)  $ is called tempered, if there exists some constant $\sigma$ such
that
\begin{equation}
\left\vert G\left(  x\right)  \right\vert <\left\vert x\right\vert ^{\sigma
};\qquad x\rightarrow\pm\infty.\label{cond_t}%
\end{equation}
Fourier transform of any tempered distribution is a tempered distribution as
well\cite{gel-shil,vladimirov,bremermann}.

The standard procedure of transition from Fourier series to Fourier integral
is based on infinite increasing of $\tau$ (see e.g. \cite{titch,Schwartz}).
Usually one starts from locally summable function $\digamma\left(  x\right)  $
and constructs a truncated function $\overline{F}\left(  x\right)  $ as
\begin{equation}
\overline{F}\left(  x\right)  \equiv\left\{
\begin{array}
[c]{ccc}%
\digamma\left(  x\right)  ; & \text{if} & -\pi\tau<x<\pi\tau\\
0; & \text{if} & \left\vert x\right\vert >\pi\tau
\end{array}
\right. \label{trunc}%
\end{equation}
Then $\overline{F}\left(  x\right)  $ is extended periodically (with the
period $2\pi\tau$) on the whole real axis $x$, i.e. $\overline{F}\left(
x\right)  \rightarrow\widetilde{F}\left(  x/\tau\right)  $
where\footnote{Below the tilde always marks periodic functions with the period
$2\pi.${}}
\begin{equation}
\widetilde{F}\left(  \varphi\right)  =\widetilde{F}\left(  \varphi+2\pi\right)
\label{per-2pi}%
\end{equation}
and
\begin{equation}
\widetilde{F}\left(  x/\tau\right)  =\digamma\left(  x\right)  =\overline
{F}\left(  x\right)  ;\label{def}%
\end{equation}
for $-\pi\tau<x<\pi\tau$.

This procedure may be easily generalized for a case when $\digamma\left(
x\right)  $, $\overline{F}\left(  x\right)  $, and $\widetilde{F}\left(
x/\tau\right)  $ are distributions. It is clear, that since generally
distributions are not defined pointwise, the periodicity condition $\left(
\ref{per-2pi}\right)  $ should be interpreted as.%

\begin{equation}
\left\langle \widetilde{F}\phi_{\varphi}\right\rangle =\left\langle
\widetilde{F}\phi_{\varphi+2\pi}\right\rangle
\end{equation}
where $\phi_{\varphi}$ is arbitrary test function $\phi_{\varphi}\in S$ with
support located in the interval $\left(  \varphi-\epsilon,\varphi
+\epsilon\right)  $ with arbitrary small, but finite $\epsilon$.

It is evident that the periodic distribution $\widetilde{F}\left(
x/\tau\right)  $ may be obtained as the result of the cyclization procedure
\cite{me I} defined as%
\begin{equation}
\widetilde{F}\left(  x/\tau\right)  =\widehat{\Sigma}\overline{F}\left(
x\right)  =\sum_{m=-\infty}^{\infty}\overline{F}\left(  x+2\pi m\tau\right)
\label{per-id}%
\end{equation}

Periodic distribution may be represented as the Fourier series
\begin{equation}
\widetilde{F}\left(  x/\tau\right)  =\sum_{n=-\infty}^{\infty}F_{n}%
\exp\left\{  inx/\tau\right\} \label{F-s-tau}%
\end{equation}
where Fourier coefficients $F_{n}$ are given by the standard expression
\begin{equation}
F_{n}=\frac{1}{2\pi\tau}\int_{-\pi\tau}^{\pi\tau}\widetilde{F}\left(
x/\tau\right)  \exp\left\{  -inx/\tau\right\}  dx\label{p-fc}%
\end{equation}
Fourier coefficients can be also computed as%
\begin{equation}
F_{n}=\frac{1}{2\pi\tau}\int_{-\pi\tau}^{\pi\tau}\overline{F}\left(  x\right)
\exp\left\{  -inx/\tau\right\}  dx,
\end{equation}
Although correspondence between $F_{n}$ and $\widetilde{F}\left(
x/\tau\right)  $ is biunique, there are no biunique correspondence between
$F_{n}$ and $\overline{F}\left(  x\right)  $ in a general case (see Appendix I).

Lastly, we wish to stress that all mentioned error and ambiguity problems are
of no particular interest, if the considered transition is used only to define
Fourier integral, as it is e.g. in \cite{titch,Schwartz}. When such definition
is done, even for some restricted class of functions, further extension of
Fourier transform on a wide class of functions and even on distributions needs
no reference to the original Fourier series. Bijection is needed only for the
function and its transform, given by the Fourier integral. However, there are
cases when the transition from Fourier series to Fourier integral is of
particular interest, e.g. when $\tau\rightarrow\infty$ corresponds to the
lifting of regularization. Hence the degree of ambiguity in such transition
deserves a more detailed consideration.

\section{Transition from Fourier series to Fourier integral}

After \cite{Schwartz} we split the region of summation $-\infty<n<\infty$ into
'bursts', i.e. intervals $\tau k-\tau\delta k/2<n<\tau k+\tau\delta k/2$ and
approximately\footnote{Although the term 'approximately' isn't specified in
\cite{Schwartz}, one may interpret it so that the approximate equality is
assumed to become an exact one with the vanishing $\delta k$ and $1/\tau$.}
replace the sum by its average value in each interval%
\begin{equation}
\frac{1}{\delta k}\sum_{n=\tau k-\tau\delta k/2}^{\tau k+\tau\delta k/2}%
\exp\left\{  inx/\tau\right\}  F_{n}\simeq\tau\left[  \exp\left\{
inx/\tau\right\}  F_{n}\right]  _{n=\tau k}=\exp\left\{  ikx\right\}  \tau
F_{k\tau}\label{smoo}%
\end{equation}

Some details of partition into 'bursts' may be clarified, if we note that
after formal change $n\rightarrow k\tau$ we get from $\left(  \ref{p-fc}%
\right)  $
\begin{equation}
\frac{1}{2\pi}\int_{-\pi\tau}^{\pi\tau}\widetilde{F}\left(  x/\tau\right)
\exp\left\{  -ikx\right\}  dx=f_{k}\tau\delta k\label{def-f2}%
\end{equation}
with definition
\begin{equation}
f_{k}\equiv\tau F_{k\tau}\label{def-f}%
\end{equation}
It is assumed in \cite{Schwartz}, that when proceeding to limit $\tau
\rightarrow\infty$, the expression $\left(  \ref{def-f2}\right)  $ was
converted into the inverse Fourier transform
\begin{equation}
\frac{1}{2\pi}\lim_{\tau\rightarrow\infty}\int_{-\pi\tau}^{\pi\tau}%
\widetilde{F}\left(  x/\tau\right)  \exp\left\{  -ikx\right\}  dx=f_{k}%
\label{def-fk}%
\end{equation}
From $\left(  \ref{def-f2}\right)  $ we conclude, that it may happen only if
for large enough $\tau$, we impose the condition%
\begin{equation}
\tau\delta k=1
\end{equation}
it means that 'bursts' are defined as $\tau k-1/2<n<\tau k+1/2$, so that each
'burst' contains only the sole term $\exp\left\{  ikx\right\}  F_{k\tau}$,
where $k\tau$ is still an integer number. Relation $\left(  \ref{def-f}%
\right)  $ is nothing but a new definition so it gives nothing new concerning
the conditions under which right-hand member in $\left(  \ref{smoo}\right)  $
may be treated as an integrand of the Fourier integral in variable $k$.

For lack of something better we suggest to \textit{claim} the fulfillment of
approximate relation\footnote{The error of such approximation will be
estimated later.}%
\begin{equation}
F_{n}\exp\left\{  in\varphi\right\}  \simeq\int_{n+\epsilon-1}^{n+\epsilon
}F_{t}\exp\left\{  it\varphi\right\}  dt;\quad0<\epsilon<1\label{smooth_2}%
\end{equation}
which directly provides the desired result%
\begin{equation}
\sum_{n=-\infty}^{\infty}F_{n}\exp\left\{  in\varphi\right\}  \simeq
\int_{-\infty}^{\infty}F_{t}\exp\left\{  it\varphi\right\}  dt=\int_{-\infty
}^{\infty}\tau F_{k\tau}\exp\left\{  ikx\right\}  dk
\end{equation}
or%
\begin{equation}
\widetilde{F}\left(  x/\tau\right)  \simeq\int_{-\infty}^{\infty}f_{k}%
\exp\left\{  ikx\right\}  dk\label{appr}%
\end{equation}

If we define
\begin{equation}
F\left(  x\right)  \equiv\int_{-\infty}^{\infty}f_{k}\exp\left\{  ixk\right\}
dk\label{F_int}%
\end{equation}
then $\left(  \ref{appr}\right)  $ may be rewritten as%
\begin{equation}
\widetilde{F}\left(  x/\tau\right)  \simeq F\left(  x\right) \label{F-In}%
\end{equation}

With the definition $\left(  \ref{def-f}\right)  $ the approximate relation
$\left(  \ref{smooth_2}\right)  $ acquires the form
\begin{equation}
f_{k}\exp\left\{  ikx\right\}  \simeq\frac{1}{\delta k}\int_{k-\left(
1-\epsilon\right)  \delta k}^{k+\epsilon\delta k}f_{t}\exp\left\{
ixt\right\}  dt\label{smooth}%
\end{equation}
Relation $\left(  \ref{smooth}\right)  $ resembles the result of
the\textit{\ mean value theorem}, however\textit{\ }$f_{t}$ should not
obligatory obey the restrictive conditions of such\textit{\ }%
theorem\textit{\ }because $\left(  \ref{smooth}\right)  $ must be satisfied
only for infinitesimal $\delta k$ and only approximately.

Making allowance for $\left(  \ref{p-fc}\right)  $, the inverse Fourier
transform may be written with the same accuracy as%

\begin{equation}
f_{k}=\frac{1}{2\pi}\int_{-\pi\tau}^{\pi\tau}F\left(  x\right)  \exp\left\{
-ikx\right\}  dx\label{def-fk-appr}%
\end{equation}

In other words, \textit{ordinary} transition may be done by the formal
substitution
\begin{equation}
n\rightarrow k\tau;\qquad\sum_{n=-\infty}^{\infty}\rightarrow\tau\int
_{-\infty}^{\infty}dk;\qquad F_{n}\rightarrow F_{k\tau}\equiv\frac{1}{\tau
}f_{k}.\label{formal}%
\end{equation}
In this paper we try to clarify under what conditions substitution $\left(
\ref{formal}\right)  $ leads to%
\begin{equation}
\lim_{\tau\rightarrow\infty}\widetilde{F}\left(  x/\tau\right)  =F\left(
x\right) \label{exa-trans}%
\end{equation}
which we treat as an exact result.

It is clear that nontrivial result appears after proceeding to limit
$\tau\rightarrow\infty$ in $\left(  \ref{def-f}\right)  $ and $\left(
\ref{exa-trans}\right)  $, only if $F_{t}$ and $\widetilde{F}\left(
x/\tau\right)  $ have specific properties. Some of them are discussed in
Appendix II.

\section{Ambiguity in choosing of extension}

In the distribution theory Fourier series exist even for such distributions
that obey neither Riemann-Lebesgue lemma nor the Dirichlet condition. As it is
proved in \cite{Schwartz}, $F_{n}$ exist and are called Fourier coefficients,
if $\widetilde{F}\left(  \varphi\right)  $ is locally summable (i.e. at any
finite interval). Distribution theory allows to sum series in which $F_{n}$ is
a tempered function of discrete argument $n$. Therefore, an appropriate
tempered distribution may be taken, as an extension $F_{n}$ for noninteger
$n$. For instance, if Fourier coefficients are given by
\begin{equation}
F_{n}=\left\{
\begin{array}
[c]{ccc}%
n^{\lambda}; & \text{if} & n>0\\
0; & \text{if} & n\leq0
\end{array}
\right.  ;\quad\lambda=\operatorname{Re}\lambda=const
\end{equation}
straightforward extension $n\rightarrow t$ gives the distribution
\begin{equation}
F_{t}=\left\{
\begin{array}
[c]{ccc}%
t^{\lambda}; & \text{if} & t>0\\
0; & \text{if} & t\leq0
\end{array}
\right.  \equiv t_{+}^{\lambda}.\label{stf}%
\end{equation}
If extension $F_{t}^{\left[  1\right]  }$ is allowable, then $F_{t}^{\left[
2\right]  }$ is allowable too under the condition that $F_{t}^{\left[
1\right]  }-F_{t}^{\left[  2\right]  }\equiv\omega_{t}$ turns into zero for
all integer $t=n$, i.e. $\omega_{t}\in\Omega$ \cite{me I}. However, when a
class of the sought distribution is specified, the choice of extension may be
essentially restricted. If, for example, it is required that the target
distribution must have continuous derivative up to $m$-th order in some
specified area (e.g. in the vicinity $t=0$), then straightforward extension
$\left(  \ref{stf}\right)  $ is unsuitable if $m>\lambda$.

There exists an extension of $F_{n}$ with exceptional analytical properties,
namely the entire function
\begin{equation}
F_{t}=\frac{1}{2\pi\tau}\int_{-\pi\tau}^{\pi\tau}\widetilde{F}\left(
x/\tau\right)  \exp\left\{  -itx/\tau\right\}  dx\label{ent-ext}%
\end{equation}
which may be obtained from expression $\left(  \ref{p-fc}\right)  $. When
according to problem situation the extension $\left(  \ref{ent-ext}\right)  $
is prescribed, then even if $F_{t}$ breaks the conditions of Carlson theorem
(see e.g.\cite{bieber}), it may be shown that such extension \cite{me I} is unique.

On the other hand, if $\widetilde{F}\left(  x/\tau\right)  $ is known,
$F\left(  x\right)  $ may be simply computed with $\left(  \ref{exa-trans}%
\right)  $. In many interesting and more common situations technical
difficulties prevent summarizing series in $\left(  \ref{F-s-tau}\right)  $.

Formally, an extension $\left(  \ref{ent-ext}\right)  $ may be reconstructed
from any other, without direct computation of series in $\left(
\ref{F-s-tau}\right)  $. Indeed, the substitution of $\left(  \ref{F-s-tau}%
\right)  $ in $\left(  \ref{ent-ext}\right)  $ gives\footnote{The Abel-Poisson
regularization is implied, so the change of the order of summation and
integration is legal.}:
\begin{equation}
F_{t}=\frac{1}{2\pi}\int_{-\pi}^{\pi}\exp\left\{  -it\varphi\right\}
\sum_{n=-\infty}^{\infty}F_{n}\exp\left\{  in\varphi\right\}  d\varphi
=\sum_{n=-\infty}^{\infty}F_{n}\frac{\sin\pi\left(  n-t\right)  }{\pi\left(
n-t\right)  }\label{entire}%
\end{equation}
where $F_{t}$ is an entire function.\ Unfortunately, the formal expression
$\left(  \ref{entire}\right)  $ adds nothing to computation of $f_{k}$ or
$F\left(  x\right)  $, since $\left(  \ref{formal}\right)  $ with
(\cite{brych-prud} 5$\left(  58\right)  $) gives
\begin{equation}
f_{k}=\frac{1}{\pi}\sum_{n=-\infty}^{\infty}F_{n}\frac{\sin\pi\left(
k-\frac{n}{\tau}\right)  \tau}{k-\frac{n}{\tau}}=\sum_{n=-\infty}^{\infty
}F_{n}\delta\left(  k-\frac{n}{\tau}\right) \label{entire 2}%
\end{equation}
and Fourier transform of $\left(  \ref{entire 2}\right)  $ brings us back to
$\left(  \ref{F-s-tau}\right)  $.

If we take an exact summation of series in $\left(  \ref{entire}\right)  $ we,
undoubtedly, obtain a practicable expression. Unfortunately, as a rule the sum
computation in $\left(  \ref{entire}\right)  $ is as difficult as the
summation of series in $\left(  \ref{F-s-tau}\right)  $, so we consider the
simple example
\begin{equation}
F_{n}=\frac{i\sinh\nu\pi}{\pi}\frac{\cos\pi n}{n+i\nu}\label{E1}%
\end{equation}
where $\nu$ is positive noninteger number. From $\left(  \ref{entire}\right)
$ we find that the corresponding entire function is%
\begin{equation}
F_{t}=\frac{\sin\pi\left(  t+i\nu\right)  }{\pi\left(  t+i\nu\right)
}\label{E2}%
\end{equation}

As it should be, the difference of $F_{t}$ obtained from $\left(
\ref{E1}\right)  $ by the straightforward extension $n\rightarrow t$ and
$F_{t}$ from $\left(  \ref{E2}\right)  $%
\begin{equation}
\omega_{t}^{\left(  \nu\right)  }=\frac{\cosh\nu\pi}{\pi}\frac{\sin t\pi
}{t+i\nu}\in\Omega\label{E3}%
\end{equation}
For sure, Fourier series either with the coefficients given in $\left(
\ref{E1}\right)  $ or those given in $\left(  \ref{E2}\right)  $ converge to
the same periodic function $\exp\left\{  \nu\widetilde{\varphi}\right\}  $,
which is equal $\exp\left\{  \nu\varphi\right\}  $ in the interval
$-\pi<\varphi<\pi$.

\section{Error and ambiguity introduced by formal transition}

Although procedure proposed in \cite{titch,Schwartz} is entirely appropriate
to define the Fourier integral, being applied to transition from Fourier
series to Fourier integral, for a wider class of functions it introduces
appreciable difference between $F\left(  x\right)  $ obtained after formal
transition $\left(  \ref{formal}\right)  $ and an exact one: $F\left(
x\right)  =\lim_{\tau\rightarrow\infty}\widetilde{F}\left(  x/\tau\right)  $.
To estimate the error we apply the Poisson formula
\begin{equation}
\sum_{n=-\infty}^{\infty}\Phi_{n}\exp\left\{  in\varphi\right\}
=\sum_{m=-\infty}^{\infty}\int_{-\infty}^{\infty}e^{i\left(  \varphi-2\pi
m\right)  t}\Phi_{t}dt\label{poisson}%
\end{equation}
to the right-hand member in $\left(  \ref{F-s-tau}\right)  $. It gives%
\begin{equation}
\widetilde{F}\left(  x/\tau\right)  =\sum_{m=-\infty}^{\infty}\int_{-\infty
}^{\infty}F_{n}\exp\left\{  in\frac{x}{\tau}-2\pi imn\right\}  dn
\end{equation}
that with $\left(  \ref{def-f}\right)  $ may be rewritten as%
\begin{equation}
\widetilde{F}\left(  x/\tau\right)  =\sum_{n=-\infty}^{\infty}\int_{-\infty
}^{\infty}f_{k}\exp\left\{  ik\left(  x+2\pi n\tau\right)  \right\}
dk\label{poi_}%
\end{equation}
and, taking into account $\left(  \ref{F_int}\right)  $, we finally get%

\begin{equation}
\widetilde{F}\left(  x/\tau\right)  =\sum_{n=-\infty}^{\infty}F\left(  x-2\pi
n\tau\right) \label{per-FI}%
\end{equation}
Since both $\left(  \ref{def-f}\right)  $ and $\left(  \ref{F_int}\right)  $
are simply definitions and include no approximations, $\left(  \ref{per-FI}%
\right)  $ is an exact expression. Comparing it with $\left(  \ref{F-In}%
\right)  $ we see that $\widetilde{F}\left(  x/\tau\right)  $ differs from
$F\left(  x\right)  $ in $\left(  \ref{F_int}\right)  $ by
\begin{equation}
\Delta\left(  x\right)  =\sum_{n\neq0}F\left(  x-2\pi\tau n\right) \label{del}%
\end{equation}
With $\tau\rightarrow\infty$ correction $\Delta\left(  x\right)  $ must
disappear for any physical value $F\left(  x\right)  $. It is not true,
however, for auxiliary values, e.g. for gauge fields.

For instance, from%
\begin{equation}
F_{t}=\omega_{t}=\left(  1-\exp\left\{  2i\pi t\right\}  \right)  R_{t}%
\in\Omega
\end{equation}
which turns into zero\footnote{The distribution $R_{t}$ is assumed not too
singular at integer $t$ to spoil that condition.} for all integer $t$, we
obtain after the formal substitution $\left(  \ref{formal}\right)  $
\begin{equation}
f_{k}=\left(  1-\exp\left\{  i2\pi k\tau\right\}  \right)  r_{k}\label{omega}%
\end{equation}
where%
\begin{equation}
r_{k}=\tau R_{k\tau}%
\end{equation}
so Fourier transform of $\left(  \ref{omega}\right)  $ gives%
\begin{equation}
F\left(  x\right)  =r\left(  x\right)  +r\left(  x+2\pi k\tau\right)
\end{equation}
hence if
\begin{equation}
r\left(  x\right)  \equiv\int_{-\infty}^{\infty}r_{k}e^{ikx}dk
\end{equation}
decreases with $x\rightarrow\infty$, then for $\tau\rightarrow\infty$ and any
finite $x$ we get $F\left(  x\right)  =r\left(  x\right)  $, despite the fact
that corresponding Fourier series is identically zero.

Nonetheless, in a case when extension\textit{\ }$\left(  \ref{ent-ext}\right)
$ is prescribed, no ambiguity arises. Indeed, if we apply the formal
substitution to extension\textit{\ }$\left(  \ref{ent-ext}\right)  $, we get
for $\tau\rightarrow\infty$
\begin{equation}
f_{k}=\tau F_{k\tau}=\frac{1}{2\pi}\int_{-\pi\tau}^{\pi\tau}\widetilde
{F}\left(  x/\tau\right)  e^{-ikx}dx\label{ent+form}%
\end{equation}
On the other hand, from the definition $\left(  \ref{def-fk}\right)  $ we
obtain%
\begin{align}
f_{k}  & =\sum_{n=-\infty}^{\infty}\frac{1}{2\pi}\int_{-\pi\tau+2\pi n\tau
}^{\pi\tau+2\pi n\tau}F\left(  x\right)  e^{-ikx}dx\\
& =\frac{1}{2\pi}\int_{-\pi\tau}^{\pi\tau}\sum_{n=-\infty}^{\infty}F\left(
x+2\pi n\tau\right)  e^{-ikx}dx\nonumber
\end{align}
that with $\left(  \ref{ent+form}\right)  $ evidently leads to the exact
result\textit{\ }$\left(  \ref{per-FI}\right)  $.

\section{Asymptotic series expansion}

As it can be seen from $\left(  \ref{poi_}\right)  $, one may estimate the
correction $\Delta\left(  x\right)  $ by computing the limit of $\exp\left\{
ik2\pi n\tau\right\}  f_{k}$ for $\tau\rightarrow\infty$. In
\cite{br-shir,brych} the method of asymptotic series expansion\ of
$\exp\left\{  ik\tau\right\}  e^{i\tau k}f_{k}$ for $\tau\rightarrow\pm\infty$
was suggested. In particular, it is shown that $e^{i\tau k}f_{k}$ vanish in
such limit, if $f_{k}$ is analytical in the vicinity of $k=0$. In our case it
means the vanishing of $\Delta\left(  x\right)  $. An exhaustive investigation
of the power type distributions, i.e. distributions of a form $f\left(
x\right)  \sim\left(  x\pm i0\right)  ^{\lambda}\ln^{m}\left(  x\pm i0\right)
$\ has been undertaken in \cite{brych,brych-prud}. In \cite{me A} this method
was extended over a family of distributions and the case when the
regularization term in $f_{k}$\ depends on $\tau$ was considered. It was shown
that%
\begin{equation}
e^{i\tau k}f_{k}=e^{-i\tau x}\left(  f_{k+i\nu/\tau}^{\left(  +\right)
}-f_{k-i\nu/\tau}^{\left(  -\right)  }\right)  =\sum_{n=0}^{\infty}%
C_{n}\left(  \tau,\nu\right)  \delta^{\left(  n\right)  }\left(  k\right)
\label{exp-nu 2}%
\end{equation}
where%

\begin{equation}
f_{k}^{\left(  +\right)  }=\int_{0}^{\infty}F\left(  t\right)  \exp\left\{
itk\right\}  dt;\quad f_{k}^{\left(  -\right)  }=-\int_{-\infty}^{0}F\left(
t\right)  \exp\left\{  itk\right\}  dt\label{inv F+}%
\end{equation}
and%
\begin{equation}
C_{n}\left(  \tau,\nu\right)  \equiv\frac{i^{-n}}{2\pi}\sum_{m=0}^{n}%
\frac{F^{\left(  n-m\right)  }\left(  \tau\right)  }{\left(  n-m\right)
!}\frac{\left(  -\nu/\left\vert \tau\right\vert \right)  ^{m}}{m!}%
\end{equation}
where $F^{\left(  n\right)  }\left(  t\right)  $ is n-th derivative of the
Fourier transform of $f_{k}$

In particular, for $\lambda=-1$ we get%
\begin{equation}
\allowbreak e^{-ik\tau}\left(  k+i\frac{\nu}{\tau}\right)  ^{-1}=%
\begin{array}
[c]{ccc}%
\mp ie^{\pm\nu}\sum_{n=0}^{\infty}\nu^{n}\left\vert \tau\right\vert
^{-n}i^{-n}\delta^{(n)}\left(  k\right)  & \text{if} & \tau\rightarrow
\pm\infty
\end{array}
\end{equation}
and%
\begin{equation}
\allowbreak e^{ik\tau}\left(  k+i\frac{\nu}{\tau}\right)  ^{-1}=%
\begin{array}
[c]{ccc}%
0 & \text{if} & \tau\rightarrow\pm\infty
\end{array}
\end{equation}
so returning to the example considered above, we see that after formal
substitution $\left(  \ref{formal}\right)  $ from $\left(  \ref{E1}\right)
$\ we get for $\tau\rightarrow\infty$%
\begin{equation}
F_{n}=\frac{i\sinh\nu\pi}{\pi}\frac{\cos n\pi}{n+i\nu}\rightarrow\tau
F_{k\tau}=f_{k}^{\left(  c\right)  }=\sinh\left(  \nu\pi\right)  e^{-\pi\nu
}\delta\left(  k\right)  +O\left(  1/\tau\right)
\end{equation}
and from $\left(  \ref{E3}\right)  $ we obtain
\begin{equation}
\omega_{n}^{\left(  \nu\right)  }=\frac{\cosh\nu\pi}{\pi}\frac{\sin n\pi
}{n+i\nu}\rightarrow\tau\omega_{k\tau}^{\left(  \nu\right)  }=f_{k}^{\left(
s\right)  }=\cosh\left(  \nu\pi\right)  e^{-\pi\nu}\delta\left(  k\right)
+O\left(  1/\tau\right)
\end{equation}
so%
\begin{equation}
f_{k}=f_{k}^{\left(  c\right)  }+f_{k}^{\left(  s\right)  }=\delta\left(
k\right)
\end{equation}
and, consequently,
\begin{equation}
F\left(  x\right)  =1
\end{equation}
that, indeed, coincides with the exact result%

\begin{equation}
F\left(  x\right)  =\lim_{\tau\rightarrow\infty}\widetilde{F}\left(
x/\tau\right)  =\lim_{\tau\rightarrow\infty}\exp\left\{  i\nu\widetilde
{x/\tau}\right\}  =1.
\end{equation}

\section{Conclusions}

The transition from Fourier series to Fourier integrals is considered and
errors introduced by the formal substitution $\left(  \ref{formal}\right)  $
are estimated. Also the ambiguity caused by transition from discrete function
to continuous one is considered. Conditions under which such ambiguity
disappears are suggested.

\section{Appendix I.}

As it is shown in \cite{me I}, the distribution%
\begin{equation}
\Delta\overline{F}\left(  x\right)  =\rho\left(  \frac{x}{\tau}-\pi\right)
-\rho\left(  \frac{x}{\tau}+\pi\right)  \in\Omega^{\prime}%
\end{equation}
have only trivial periodical extension
\begin{equation}
\Delta\widetilde{F}\left(  x/\tau\right)  \triangleq\widehat{\Sigma}\left(
\Delta\overline{F}\left(  x\right)  \right)  =0,\label{del-fin}%
\end{equation}
so both $\overline{F}\left(  x\right)  $ and $\overline{F}_{\rho}\left(
x\right)  \equiv\overline{F}\left(  x\right)  +\Delta\overline{F}\left(
x\right)  $ after the cyclization procedure $\left(  \ref{per-id}\right)  $
give the same $\widetilde{F}\left(  x/\tau\right)  $. It also means that
inverse procedure $\overline{F}\left(  x\right)  \triangleq\widehat{\Sigma
}^{-1}\widetilde{F}\left(  x/\tau\right)  $ is ambiguous, i.e. $\overline
{F}\left(  x\right)  $ cannot be reconstructed uniquely for a given
$\widetilde{F}\left(  x/\tau\right)  $. Since $\overline{F}_{\rho}\left(
x\right)  $ should vanish outside $-\pi\tau\leq x\leq\pi\tau$, distributions
$\rho\left(  \frac{x}{\tau}\pm\pi\right)  $ may differ from zero only at the
points $x=\mp\pi\tau$, i.e. $\rho\left(  \frac{x}{\tau}\pm\pi\right)  $ have
point support. As it is known (see e.g. \cite{gel-shil,bremermann}),
distribution with point support is finite linear combinations of Dirac
$\delta$-functions and its derivatives, so one may choose
\begin{equation}
\Delta\overline{F}\left(  x\right)  =\sum_{m=0}^{M}c_{m}\left[  \delta
^{\left(  m\right)  }\left(  \frac{x}{\tau}-\pi\right)  -\delta^{\left(
m\right)  }\left(  \frac{x}{\tau}+\pi\right)  \right]  ,\label{exa-delta}%
\end{equation}
with arbitrary finite constants $c_{m}$ and $M$.

As it can be seen from $\left(  \ref{del-fin}\right)  $ no contribution may
come from $\Delta\overline{F}\left(  x\right)  $ to Fourier coefficients, in
other words%
\begin{equation}
\Delta\overline{F}_{t}\equiv\frac{1}{2\pi\tau}\int_{-\pi\tau}^{\pi\tau}%
\Delta\overline{F}\left(  x\right)  e^{-itx/\tau}dx
\end{equation}
must turn into zero for all integer $t=n$. To guarantee this, we must remove
some uncertainty in integration $\delta^{\left(  m\right)  }\left(  \frac
{x}{\tau}\pm\pi\right)  $ on $\left[  -\pi\tau,\pi\tau\right]  $. Recall that
Dirac $\delta$-function is defined as the functional$\int_{-\infty}^{\infty
}\phi\left(  y\right)  \delta\left(  y\right)  dy=\phi\left(  0\right)  $, for
any test function $\phi\left(  y\right)  $ continuous in the vicinity $y=0$.
However, there exists no such $\phi\left(  y\right)  $ with support located in
the interval $0\leq y<\infty$, i.e.
\begin{equation}
\left\langle \phi,\delta\right\rangle =\int_{-\infty}^{\infty}\phi\left(
y\right)  \delta\left(  y\right)  dy=\int_{0}^{\infty}\phi\left(  y\right)
\delta\left(  y\right)  dy
\end{equation}
with $\phi\left(  0\right)  \neq0$. Certainly $\left\langle \phi
,\delta\right\rangle =0$ does not look winning at all. Even if we put
$\int_{-y}^{\infty}\delta\left(  y^{\prime}\right)  dy^{\prime}=\theta\left(
y\right)  $, then
\begin{equation}
\Delta\overline{F}_{t}=\lim_{\varepsilon\rightarrow0}\left[  e^{i\pi t}%
\theta\left(  \varepsilon\right)  -e^{-i\pi t}\theta\left(  -\varepsilon
\right)  \right]  \frac{1}{2\pi}\sum_{m=0}^{M}\left(  it\right)  ^{m}c_{m}%
\end{equation}
so we only shifted the trouble to $\theta\left(  x\right)  $, which is
undefined at $x=0$. To be more exact, value $\theta\left(  0\right)  $ may be
chosen arbitrarily.

Indeed, let as consider some function $D\left(  x\right)  $ which possesses
arbitrary values on a countable set of $x$ and is equal to zero on a
complement of such set. We assume also that $D\left(  x\right)  $ does not
include linear combinations of Dirac $\delta$-functions and its derivatives.
In this case the functional $\left\langle D,\phi\right\rangle =0$ for any of
the test functions $\phi$. The distribution theory expressly makes no
distinction between distributions, which differ by $D\left(  x\right)  $, if
$D\left(  x\right)  $ satisfies the above condition. In other words, on a
countable set of points one may arbitrarily choose values of any distribution,
if at these points such distribution can not be presented as Dirac $\delta
$-functions and its derivatives.

In particular, one may choose so called 'symmetric' definition
\begin{equation}
\theta\left(  x\right)  =\lim_{\varepsilon\rightarrow0}\left(  \frac
{\theta\left(  x+\varepsilon\right)  +\theta\left(  x-\varepsilon\right)  }%
{2}\right)  ,
\end{equation}
that gives $\theta\left(  0\right)  =1/2$. The 'symmetric' definition is
especially convenient, if distribution is represented as the Fourier series,
because, according to Fej\'{e}r theorem (see e.g. \cite{whit-wats}), if
Fourier series\ is convergent\ at $\varphi+0$\ and $\varphi-0,$ it converges
to $\widetilde{F}\left(  \varphi\right)  =\left(  \widetilde{F}\left(
\varphi+0\right)  +\widetilde{F}\left(  \varphi-0\right)  \right)  /2$. The
'symmetric' definition allows to get for $\Delta\overline{F}_{t}$ an
acceptable result.
\begin{equation}
\Delta\overline{F}_{t}=i\frac{\sin\left(  \pi t\right)  }{\pi}\sum_{m=0}%
^{M}\left(  it\right)  ^{m}c_{m}%
\end{equation}
Nevertheless, that or another choice is merely a matter of convenience,
because this problem lies outside the scope of the distribution theory. For
example, in \cite{brych-prud}8.1$\left(  732\right)  $ another definition is
taken
\begin{equation}
\int_{0}^{\infty}e^{-px}\delta\left(  x\right)  dx=\int_{0}^{\infty}%
\delta\left(  x\right)  dx=1
\end{equation}
i.e. $\theta\left(  0\right)  =1$.

\section{Appendix II. Some remarks about analytical properties of $F_{t}$ and
$\widetilde{F}\left(  x/\tau\right)  $}

As it already mentioned, after proceeding to limit $\tau\rightarrow\infty$ the
nontrivial result may appear, only if $F_{t}$ and $\widetilde{F}\left(
x/\tau\right)  $ have specific properties. Indeed, if $\widetilde{F}\left(
x/\tau\right)  $ depends on $x$ and $\tau$ only through $x/\tau$, then
$\lim_{\tau\rightarrow\infty}\widetilde{F}\left(  x/\tau\right)  =$
$\widetilde{F}\left(  0\right)  $ for any $x$ where such limit exists, unless
$x $ is the infinite point. In particular for any finite Fourier series, one
may write%
\begin{equation}
\widetilde{F}\left(  x/\tau\right)  =\sum_{n=-N_{-}}^{N_{+}}F_{n}\exp\left\{
inx/\tau\right\}  \rightarrow\sum_{n=-N_{-}}^{N_{+}}F_{n}=\widetilde{F}\left(
0\right)  =const
\end{equation}
for any fixed $N_{\pm}$.

In proceeding to limit $\tau\rightarrow\infty$\ the positions of singularity
of $\tau F_{k\tau}$\ and $\widetilde{F}\left(  x/\tau\right)  $\ are
essentially rearranged. For instance, if all singularities of $F_{t}$ are
located in the finite area $\left\vert t-t_{0}\right\vert <R$, where $t_{0}$
and $R$ may be arbitrarily large, but finite, then for $\tau\rightarrow\infty$
all corresponding singularities of $f_{k}$ are moved to the infinitely small
vicinity of $k=0$. Indeed, in our case $F_{t}$ may be expanded in Laurent
series%
\begin{equation}
F_{t}=\sum_{m=1}^{\infty}\frac{\lambda_{m}}{\left(  t-t_{0}\right)  ^{m}%
}\label{L}%
\end{equation}
for any $t$ which obeys $\left\vert t-t_{0}\right\vert >R$. Therefore, after
the formal substitution $\left(  \ref{formal}\right)  $ we get%
\begin{align}
F_{t}  & \rightarrow f_{k}=\lim_{\tau\rightarrow\infty}\tau F_{k\tau}%
=\lim_{\tau\rightarrow\infty}\sum_{m=1}^{\infty}\frac{\tau^{1-m}\lambda_{m}%
}{\left(  k-t_{0}/\tau\right)  ^{m}}\\
& =\frac{\lambda_{1}}{k-i\varepsilon\operatorname*{signum}\left(
\operatorname{Im}t_{0}\right)  \vspace{6pt}}\nonumber
\end{align}

Less trivial and more realistic result may be obtained only when $\tau
$-dependence of $\lambda_{m}$ is allowed. We exclude the case when
$\lambda_{m}$ increase faster than $\tau^{m-1}$ with $\tau\rightarrow\infty$,
because it makes $f_{k}$ infinite for all $k$. Therefore, either $\lambda
_{m}\rightarrow c_{m}\tau^{m-1}$ or corresponding terms vanish. Thereby, we
may write%
\begin{equation}
f_{k}=\lim_{\tau\rightarrow\infty}\sum_{m=1}^{\infty}\frac{c_{m}}{\left(
k-t_{0}/\tau\right)  ^{m}}=\lim_{\varepsilon\rightarrow+0}\sum_{m=1}^{\infty
}\frac{c_{m}}{\left(  k-i\varepsilon\operatorname*{signum}\left(
\operatorname{Im}t_{0}\right)  \right)  ^{m}}.\label{L-ex}%
\end{equation}
So if series $\left(  \ref{L-ex}\right)  $ converges, $f_{k}$ became an
analytic function in a whole complex plane $k$, except the point $k=0$.

Let now $\widetilde{F}\left(  \varphi\right)  $ be regular in the area
$\left\vert \varphi\right\vert \leq c$, where $c$ is arbitrary small, but
finite, then $\widetilde{F}\left(  x/\tau\right)  $, and consequently
$F\left(  x\right)  $, may have singularity only at $\left\vert x\right\vert
>c\tau$, so after $\tau\rightarrow\infty$ distribution $\widetilde{F}\left(
x/\tau\right)  $ converts into entire function. When $\widetilde{F}\left(
\varphi\right)  $ depends not only on $\varphi$, but on $\tau$ as well, i.e.
$\widetilde{F}=\widetilde{F}\left(  \varphi,\tau\right)  $, and if
$\widetilde{F}\left(  \varphi,\tau\right)  $ is analytical in a ring
$r_{-}<\left\vert \varphi\right\vert <r_{+}$, i.e.
\begin{equation}
\widetilde{F}\left(  \varphi,\tau\right)  =\sum_{n=-\infty}^{\infty}%
\widetilde{F}_{n}\left(  0,\tau\right)  \varphi^{-n-1}%
\end{equation}
with $r_{\pm}=\overline{\lim}_{n\rightarrow\pm\infty}\left\vert \widetilde
{F}_{n}\left(  0,\tau\right)  \right\vert ^{-\frac{1}{n}}$, such
$\widetilde{F}\left(  \varphi,\tau\right)  $ may have nontrivial behavior for
infinitely increasing period under the conditions which may be easily found.
Indeed, within such ring $\widetilde{F}\left(  \varphi,\tau\right)  $ may be
presented as Laurent series and for a reason similar to the one mentioned
above we obtain for $\tau\rightarrow\infty$
\begin{equation}
\widetilde{F}_{n}\left(  0,\tau\right)  \rightarrow\tau^{n}\Phi_{n};\qquad
\Phi_{n}=const
\end{equation}
so the distribution
\begin{equation}
\Phi\left(  x\right)  \equiv\lim_{\tau\rightarrow\infty}\widetilde{F}\left(
\frac{x}{\tau},\tau\right)  =\sum_{n=-\infty}^{\infty}\Phi_{n}x^{-n-1}%
\end{equation}
is analytic in a ring $R_{-}<\left\vert x\right\vert <R_{+}$ with $R_{\pm
}=\overline{\lim}_{n\rightarrow\pm\infty}\left\vert \Phi_{n}\right\vert
^{-\frac{1}{n}}$.

\end{document}